\documentclass[
reprint,
superscriptaddress,
showpacs,
amsmath,amssymb,
aps,prl,
]{revtex4-1}
\usepackage{amssymb}
\usepackage{amsmath}
\usepackage{booktabs}
\usepackage{tabularx}
\usepackage{braket}
\usepackage{float}
\usepackage{graphicx}
\usepackage{dcolumn}
\usepackage{bm}
\usepackage{subfigure}
\usepackage{color}

\usepackage{epstopdf, epsfig}
\usepackage{blindtext}
\usepackage{longtable}
\usepackage{amsmath}
\usepackage{mathtools}
\usepackage{enumitem}
\usepackage[T1]{fontenc}
\usepackage{mathptmx}

\usepackage[normalem]{ulem}
\usepackage{siunitx}
\usepackage{layouts}

\usepackage[colorlinks,citecolor = blue, linkcolor=red,hyperindex,CJKbookmarks]{hyperref}

\newcommand{\Stop}{S_\textrm{top}}
\newcommand{\Sbot}{S_\textrm{bot}}

\newcommand{\Sh}{S_\textrm{h}}

\begin{document}

\title{Ice melting in salty water: layering and non-monotonic dependence on the mean salinity} 

\author{Rui Yang}
\affiliation{Physics of Fluids Group, Max Planck Center for Complex Fluid Dynamics, and J.M.Burgers Center for Fluid Dynamics, University of Twente, P.O. Box 217, 7500 AE Enschede, The Netherlands}
\author{Christopher J. Howland}
\affiliation{Physics of Fluids Group, Max Planck Center for Complex Fluid Dynamics, and J.M.Burgers Center for Fluid Dynamics, University of Twente, P.O. Box 217, 7500 AE Enschede, The Netherlands}
\author{Hao-Ran Liu}
\affiliation{Physics of Fluids Group, Max Planck Center for Complex Fluid Dynamics, and J.M.Burgers Center for Fluid Dynamics, University of Twente, P.O. Box 217, 7500 AE Enschede, The Netherlands}
\author{Roberto Verzicco}
\affiliation{Physics of Fluids Group, Max Planck Center for Complex Fluid Dynamics, and J.M.Burgers Center for Fluid Dynamics, University of Twente, P.O. Box 217, 7500 AE Enschede, The Netherlands}
\affiliation{Dipartimento di Ingegneria Industriale, University of Rome 'Tor Vergata', Via del Politecnico 1, Roma 00133, Italy}
\affiliation{Gran Sasso Science Institute - Viale F. Crispi, 7, 67100 L'Aquila, Italy}
\author{Detlef Lohse}\email{d.lohse@utwente.nl}
\affiliation{Physics of Fluids Group, Max Planck Center for Complex Fluid Dynamics, and J.M.Burgers Center for Fluid Dynamics, University of Twente, P.O. Box 217, 7500 AE Enschede, The Netherlands}
\affiliation{Max Planck Institute for Dynamics and Self-Organisation, Am Fassberg 17, 37077 G{\"o}ttingen, Germany}

\date{\today}
\begin{abstract}
The presence of salt in ocean water strongly affects the melt rate and the shape evolution of ice, both of utmost relevance in geophysical and ocean flow and thus for the climate.
To get a better quantitative understanding of the physical mechanics at play in ice melting in salty water, we numerically investigate the lateral melting of an ice block in stably stratified saline water, using a realistic, nonlinear equation of state (EOS). The developing ice shape from our numerical results shows good agreement with the experiments and theory from 
Huppert \& Turner (\textit{J. Fluid Mech.} 100, 367 (1980)).
Furthermore, we find that the melt rate of ice depends non-monotonically on the mean ambient salinity: It first decreases for increasing salt concentration until a local minimum is attained, and then increases again.
This non-monotonic behavior of the ice melt rate is due to the competition among salinity-driven buoyancy, temperature-driven buoyancy, and salinity-induced stratification. We develop a theoretical model based on the energy balance which gives a prediction of the salt concentration for which the melt rate is minimal, and is consistent with our data. Our findings give insight into the interplay between phase transitions and double-diffusive convective flows.
\end{abstract}

\maketitle

Melting and freezing has huge relevance in various fields, with a wide range of applications in nature \& technology, including sea ice \citep{holland2006future}, phase-change materials \citep{dhaidan2015melting}, aircraft icing \citep{cao2018aircraft}, icebergs \citep{ristroph2018sculpting}, and  icy moons \citep{spencer2006cassini,kang2020spontaneous}.
Accurately quantifying the melt rate of glacial ice in the ocean is vital for constraining estimates of sea level rise under various climate change scenarios \citep{edwards_projected_2021,cenedese2022icebergs}.
The presence of salt in terrestrial (and possibly extraterrestrial) oceans introduces \emph{double-diffusive} or even \emph{multicomponent} convection driven by both temperature and salinity variations. The coupling of such flows to a moving phase boundary from a mathematical point of view constitutes a so-called Stefan problem \citep{rubinstein1971stefan}.

To better understand such highly complex systems, we consider a sufficiently simplified model problem that still contains the rich phenomenology of the turbulent flow around the melting ice observed in reality. Much previous work has focused on melting with \emph{single}-component convective flows, where the melting dynamics and convection are solely determined by temperature variations \citep{davis1984pattern,favier2019}. Extensions to this approach, using both experiments and simulations, have also considered the effects of shear \citep{hester2021aspect,couston2021topography} and rotation \citep{ravichandran2021melting,ravichandran_combined_2022} on the phase transition process, as well as the dependence on the initial conditions \citep{purseed2020bistability} and the nonlinear equation of state (EOS) \citep{wang2021growth,wang2021equilibrium,wang2021ice,yang2022abrupt}.

However, salinity significantly complicates the problem as it modifies the density and the melting point of aqueous ice. The importance of salinity on ice melting has been experimentally demonstrated by experiments by \citep{huppert1980ice} and \citep{mcconnochie2016effect}. They showed that the meltwater spreads into the liquid in a series of horizontal layers. Also, the ice forms layered structures corresponding to the flow structures. Numerical simulations have been used to study the layer structures in laterally cooled double-diffusive convection, with a temperature gradient in the horizontal direction and a salinity gradient in the vertical direction \citep{kranenborg_evolution_1998,chong2020cafe}. However, the coupling of such a flow and the melting process could up to now not be numerically modelled, due to the challenge of properly representing the salinity effect on ice melting and due to the computation time limitation. Here we will overcome these limitations to quantitatively answer how salinity affects the melt rate and shape evolution of ice.

We conduct numerical simulations of a fixed vertical ice block,  melting from the side by salt-stratified water.
The Navier-Stokes equations and the advection equations for temperature \& salinity are coupled to the phase-field for the ice-water interface, a model which is widely used for the phase boundary evolutions \citep{favier2019,hester2021aspect,couston2021topography,yang2022abrupt}. Layered structures on the melt front are observed and quantitatively agree with the experiments from \cite{huppert1980ice}. Furthermore, a non-monotonic trend is observed, where the melt rate is first reduced and then enhanced, as the salinity in the water increases. Despite the complexity of the moving boundary interaction with the turbulent flow, we provide a simple theoretical model based on an energy balance, which predicts the dependence of the minimal melt rate on salinity and temperature. 

\begin{figure}
  \centering
   \noindent\includegraphics[width=0.49\textwidth]{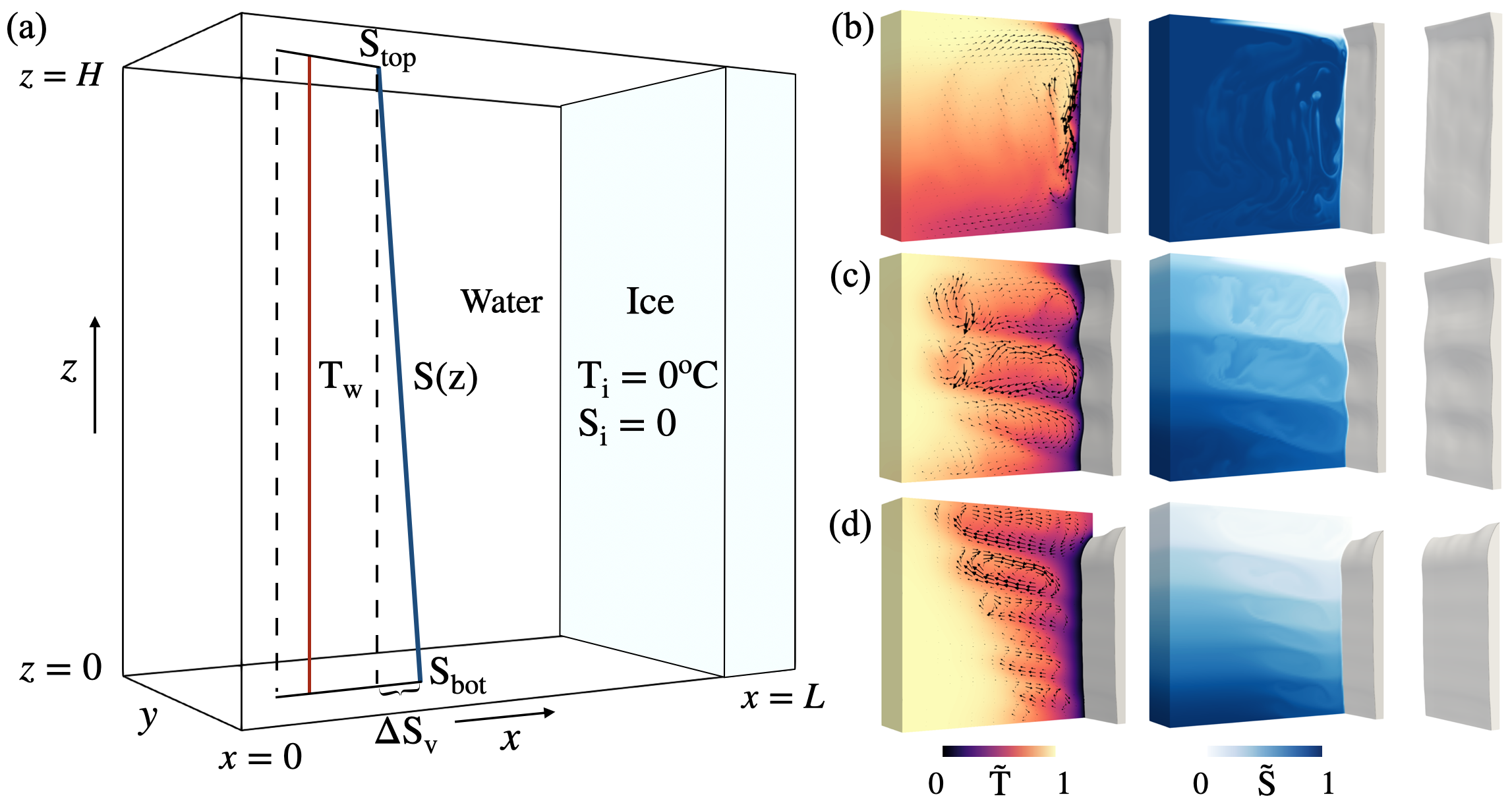}
   \caption{
   (a) An illustration of the simulation setup. Initially, the ice block is set at the right sidewall, the temperature of the water is set to be uniform as $\Delta T$ and the salinity of the water is set with a vertical gradient depending on $S_m$ and $\Delta S_v$.
   (b,c,d) Snapshots of temperature (1st), salinity field (2nd), and contour of melt front (3rd column) at $S_v=0$ (b), at $S_v=5\rm\ g/kg$ (c), and at $S_v=10\rm\ g/kg$ (d).
     }
   \label{fig:fig1}
 \end{figure}

\textbf{Numerical method and set-up}: The flow is confined to a rectangular box of height $H$ and aspect ratio $\Gamma=L_x/H=1$.
For the three-dimensional (3D) simulations, the depth-wise aspect ratio is set to $\Gamma_y = L_y/H = 0.5$.
No heat flux, no salt flux, and no-slip boundary conditions are applied on all walls. Initially, we place a vertical ice block with a thickness of $0.1H$, as shown in figure~\ref{fig:fig1}(a).
The initial temperature $T$ and salinity $S$ fields are prescribed as follows, with a linear salinity profile in the vertical ($z$) for the liquid phase:
$$
  T= \begin{cases}
    T_w, x<0.9H \\
    T_i, x\ge 0.9H
  \end{cases}
  \
  S = \begin{cases}
    \Sbot + (\Stop - \Sbot) z/H,x<0.9H \\
    0, x\ge 0.9H
  \end{cases}
$$
The initial solid temperature $T_i = \SI{0}{\celsius}$ is set to the equilibrium melting temperature.
$S_\textrm{top}$ and $S_\textrm{bot} \ge S_\textrm{bot}$ are the initial values of salinity at the top and bottom boundaries respectively.
From these initial conditions, we can define a temperature scale and two salinity scales, accounting for variations in the vertical due to stratification, and in the horizontal between the ice (which has $S_i=0$) and the mean salt concentration in the water:
\begin{align}
  \Delta T = T_w - T_i,\  \Delta S_\textrm{v} = \Stop - \Sbot,\  S_m = (\Stop + \Sbot)/2 .
\end{align}
Based on the commonly used Oberbeck-Boussinesq approximation which retains the density variation only in the buoyancy term, we use a simplified yet realistic EOS for water at atmospheric pressure \citep{roquet2015defining}, defined as
\begin{equation}
\rho'={-C_b}/{2}\left(T-T_0-c_S S\right)^2+b_0 S
\label{eq:EOS}
\end{equation}
where $\rho'=\rho - \rho_0$ is the fluid density perturbation from a reference value $\rho_0$, and the coefficients have values $C_b=\SI{0.011}{kg.m^{-3}.K^{-2}}$, $b_0=\SI{0.77}{kg.m^{-3}.(g.kg^{-1})^{-1}}$, $T_0=\SI{4}{\celsius}$, $c_S=\SI{-0.25}{K.(g.kg^{-1})^{-1}}$.

Simulations are performed using the second-order staggered finite difference code AFiD \citep{van2015pencil}, which has been extensively validated \citep{kooij2018comparison} and used to study a wide range of convection problems, including double-diffusive convection
 \citep{yang2016pnas,yang2022layering}. More details of the numerical method and governing equations are shown in the Supplementary Material. The extension of the AFiD code to include two phases approached with the phase-field method was discussed \& validated in \cite{liu2021efficient,yang2022abrupt}. 

As resulting independent dimensionless control parameters we take the thermal and solutal Rayleigh numbers, the Prandtl number, the Schmidt number, the Stefan number, and the density ratio between vertical and horizontal salinity difference $\Lambda_S$ (see also suppl. material):
\begin{equation}
\begin{split}
  Ra_T &= \frac{g C_b \Delta T^2 H^3}{2\nu\kappa_T}, 
  Ra_S = \frac{g b_0 S_m   H^3}{ \nu\kappa_S}, 
  Pr   = \frac{\nu}{\kappa_T}, \\
  Sc   &= \frac{\nu}{\kappa_S}, 
  St   = \frac{\mathcal{L}}{c_p \Delta T}, 
   \Lambda_S =\frac{\Delta S_v}{S_m} .
\end{split}
\end{equation}
Furthermore, we can define the Lewis number as the ratio of heat and salt diffusivity, as well as the density ratio between temperature and horizontal salinity difference $\Lambda_T$,  
\begin{align}
  Le &= \frac{\kappa_T}{\kappa_S} = {Sc\over Pr}, &
  \Lambda_T &=\frac{2b_0S_m}{C_b\Delta T^2} = {Ra_S \over Ra_T}.
\end{align}

Due to the large parameter space, some of the control parameters have to be fixed in order to make the study feasible. We fix $Pr=10$ and $Sc=1000$ (i.e.,  $Le=100$) as relevant values for seawater in all cases. Our simulations cover a parameter range of $\SI{10}{\celsius}\le \Delta T\le\SI{20}{\celsius}$, $0\le S_m\le\SI{15}{g.kg^{-1}}$ and $\SI{2.5}{cm}\le H\le\SI{10}{cm} $, corresponding roughly to $10^6 \le Ra_T\le 10^8$, $0 \le Ra_S\le 2.5\times10^{10}$, and $4\le St\le 8$. Unless specified, we fixed the initial temperature of the water as $T=\SI{20}{\celsius}$ and the domain height $H=\SI{5}{cm}$ (corresponding to $Ra_T=10^7$).

\textbf{Salinity effect on the structure of the melting interface}: To reveal the effect of salinity on the shape evolution and the melt rate of the ice, we begin with a qualitative description of how the melt front shape depends on the vertical salinity gradient $\Delta S_v$. In figure~\ref{fig:fig1}(b-d) typical temperature and salinity fields for three different 3D simulations are shown, where we fix $\Delta T=\SI{20}{K}$, $H=\SI{5}{cm}$, $S_m=\SI{5}{g.kg^{-1}}$, and vary $\Delta S_v$.

At relatively low or zero vertical salinity variation $\Delta S_v =0$ (figure~\ref{fig:fig1}(b)), when there is no stable stratification, salinity and temperature are mostly uniform in the bulk, and a concavely shaped ice melt front forms due to the large-scale circulation, as shown in figure~\ref{fig:fig1}(b),  right. At moderate $\Delta S_v~(=5\rm\ g/kg)$ (figure~\ref{fig:fig1}(c)), a layered structure occurs for the temperature and salinity fields, and correspondingly also for the ice melt front, similar to the experiments from \cite{huppert1980ice}. When further increasing $\Delta S_v~(=10\rm\ g/kg)$ (figure~\ref{fig:fig1}(d)), more layers appear in the liquid phase as compared to the case of $\Delta S_v=5\rm\ g/kg$, while the layer structure disappears on the melt front, since the convective flow at the interface is weakened by the stronger stable stratification.

\begin{figure}
\centering
\noindent\includegraphics[width=0.49\textwidth]{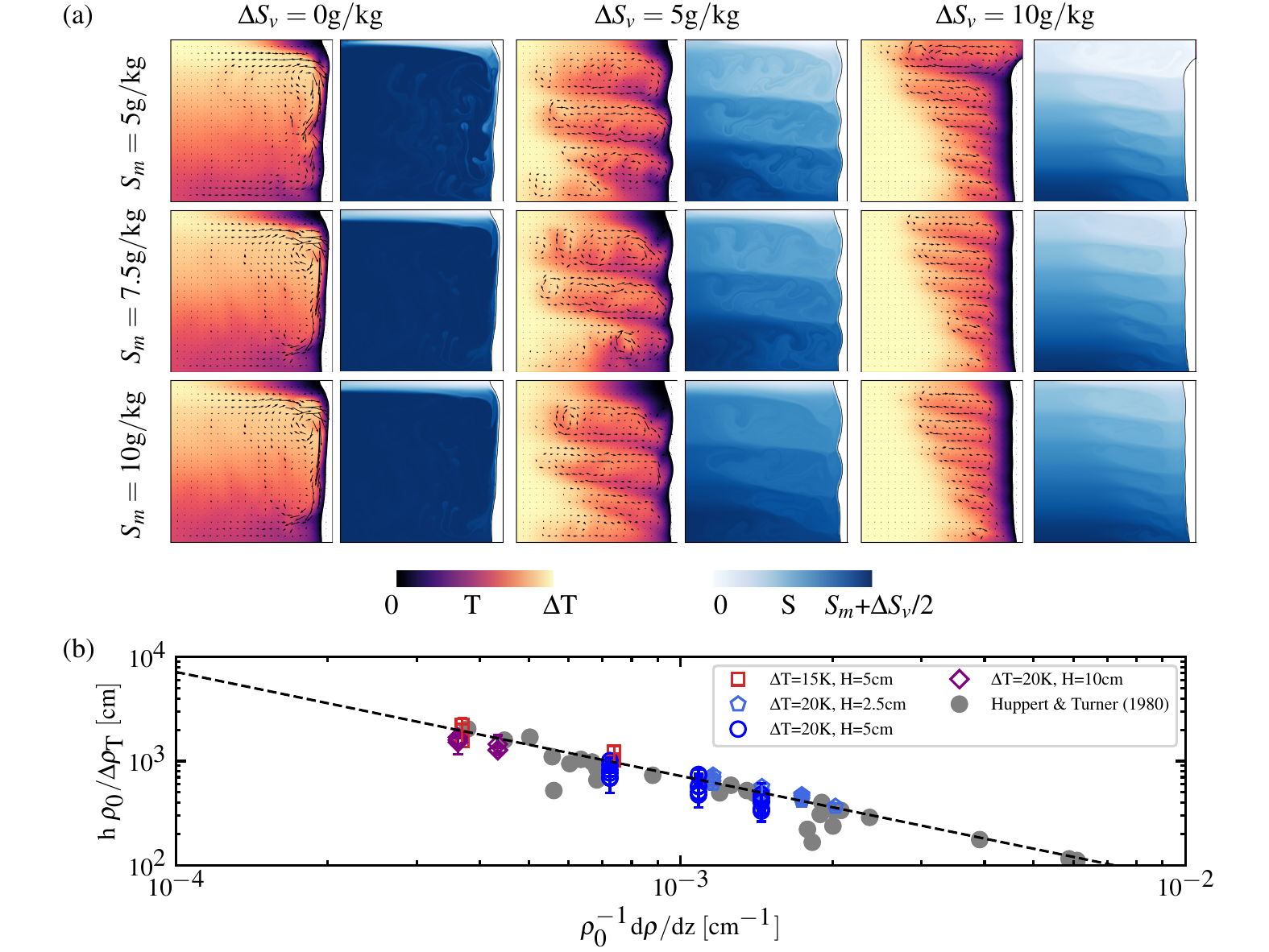}
\caption{(a) Instantaneous snapshots of temperature (left) and salinity (right) field for various $S_m$ and $\Delta S_v$. The velocity field is shown as arrows in the temperature field. (b) The layer thickness $h$ per unit density as function of the density gradient, following the same representation as in  figure 10 of
 \cite{huppert1980ice}.
 The reference density is $\rho_0$ and 
  $\Delta\rho_T=\rho(0, S_{\infty})-\rho(T_{\infty}, S_{\infty})$. The dashed line represents the theoretical result eq.~\ref{eq:prediction}.}
\label{fig:fig2}
\end{figure}

With this qualitative behaviour established, we now focus on a more detailed analysis of how both vertical and horizontal variations in salinity affect the system.
In figure~\ref{fig:fig2}(a), we present the temperature and salinity fields from 2D simulations at various horizontal \& vertical salinity variations $S_m$ and $\Delta S_v$. One can see that the flow and melt front structure mainly depends on $\Delta S_v$. Therefore, the vertical salinity gradient and horizontal temperature gradients are the main driving factors of the flow structure; consistent with the findings of \cite{huppert1980ice} and \cite{chong2020cafe}.

In all cases, a very thin boundary layer of fresh (low salinity) meltwater rises along the ice front to the top of the domain. This leads to the accumulation of cold, fresh water in the upper region which leads to a local maximum in ice thickness at the top boundary for all cases except the case with $\Delta S_v=\SI{10}{g.kg^{-1}}$ and $\Delta \Sh=\SI{5}{g.kg^{-1}}$ (see figure~\ref{fig:fig2}(a)) for which the upper region is anyway close to zero salinity, so the temperature-driven buoyancy forcing becomes stronger and even dominant close to the top boundary.

The different diffusivities of heat and salt play a significant role in the dynamics outside the thin fresh layer.
Since heat diffuses much faster than salt ($Le=100$), a region of cold, saline water is produced, which sinks due to thermal-induced buoyancy.
At $\Delta S_v = 0$, the lack of salt stratification allows this water to sink. The accumulation of cold water causes thicker ice at the bottom than in the middle, see the leftmost column in figure~\ref{fig:fig2}(a). Combined with the accumulation of fresh water at the top, this results in a concave shape of the ice, with the thinnest ice in the middle.

At $\Delta S_v=5\rm\ g/kg$, the physical explanation of the observed layer formation is as follows: after the ice starts to melt, there is an accumulation of cold water outside the thin fresh boundary layer. Being heavier than the surrounding fluid, the cold water sinks.
The surrounding fluid, however, becomes denser with depth as the local salinity also increases, and eventually the cold water reaches a neutral buoyancy level,
 thus producing a sequence of layers as seen from figure~\ref{fig:fig2}(a). In this case, vertically stacked convection rolls form. The layered convection rolls sculpt a layered pattern in the melt front since these rolls bring warm water from the bulk to the melt front and cause non-uniform heat flux at the interface.

At $\Delta S_v=\SI{10}{g/kg}$, more layers in flow structure appear as compared to the case of $\Delta S_v=\SI{5}{g/kg}$, in accordance with our physical explanation of the formation. At the top, where water is fresh, the cold meltwater descends, and at the bottom, where water is salty, the fresh meltwater ascends. Therefore, the meltwater accumulates and slows down the melting in the middle. 

We next quantitatively address the layer height $h$ of the melt front, which occurs in presence of a stable stratification.  From the analysis of \cite{huppert1980ice}, by balancing the horizontal density difference due to temperature and the vertical density gradient due to salinity, the thickness of these layers was quantified as
\begin{equation}
h=(0.65 \pm 0.06)\left[\rho(0, S_{\infty})-\rho(T_{\infty}, S_{\infty})\right]\left(\frac{\mathrm{d} \rho}{\mathrm{d} z}\right)^{-1}
\label{eq:prediction}
\end{equation}
where $\rho(T,S)$ is the fluid density at temperature $T$ and salinity $S$, $d\rho/dz$ is the ambient density stratification, the subscript $\infty$ in $T_\infty$ and $S_\infty$ relates to the mean far-field value. 

In figure~\ref{fig:fig2}(b), we plot the mean layer thickness $h$, normalised by the horizontal density difference, as a function of the stratification.
We present data both from our simulations and also from the experimental data obtained by \cite{huppert1980ice}. Our simulation results quantitatively agree with the experimental data very well, with eq.~\eqref{eq:prediction} also giving a good prediction of the layer thickness $h$. Moreover, our results of the melt front shape well match with the experiments of \cite{huppert1980ice}, which can also be regarded as a validation for our simulations of ice melting in saline water.

\begin{figure}
  \centering
  \noindent\includegraphics[width=0.5\textwidth]{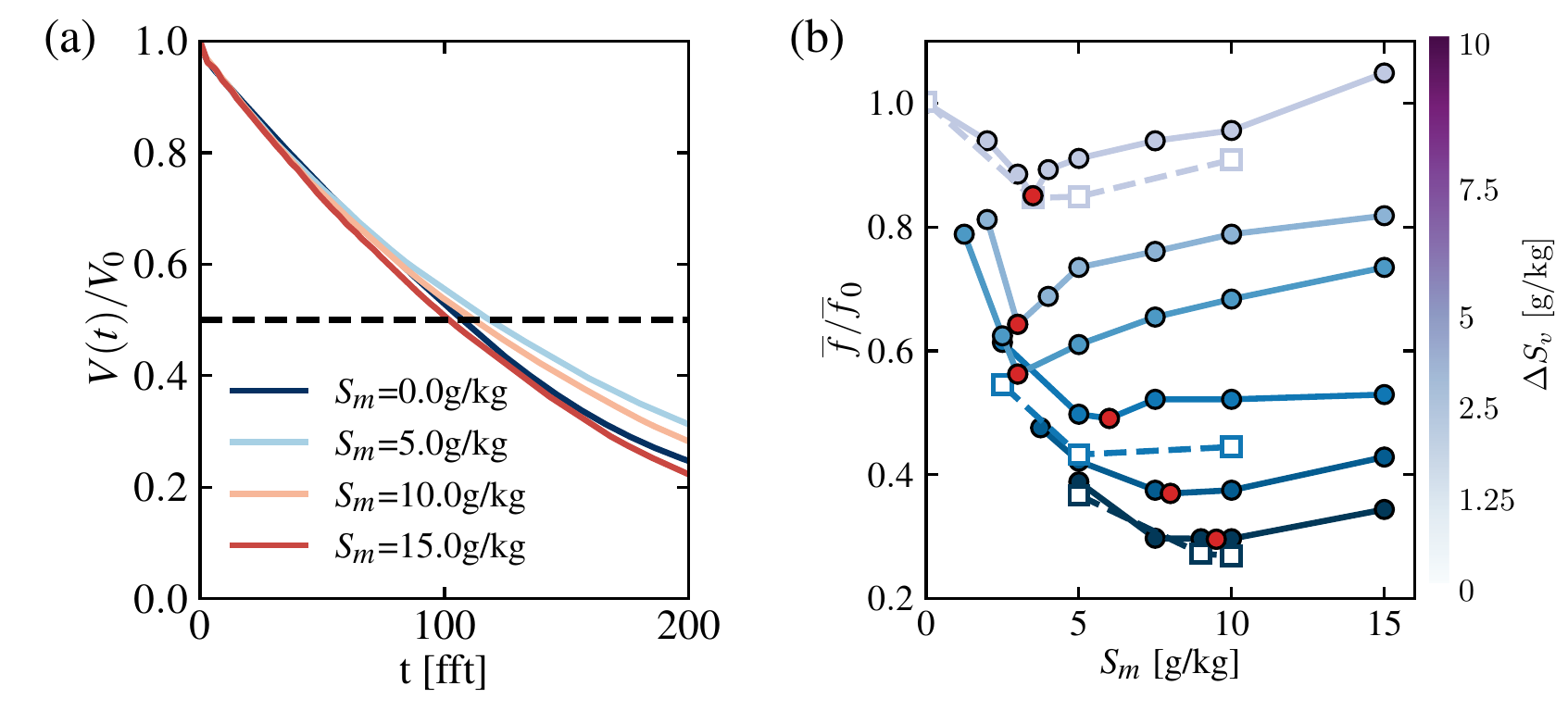}
  \caption{
    (a) The normalize volume of ice $V(t)/V_0$ as a function of time $t$ in free-fall time units. The dashed line represents the location of half of the initial ice volume. (b) The normalized melt rate $\bar{f}/\bar{f}_0$ as a function of $S_m$ for various $\Delta S_v$ (color-coded), where $\bar{f}_0$ is the melt rate without salinity. Circle data points represent 2D simulations, and square data points represent 3D simulations. The red circle data points represent the location of the minimum $\bar{f}$.
  }
  \label{fig:fig3}
\end{figure}

\textbf{A unifying view of the dependence of the average melt rate on salinity}: Our objective now is to quantify how salinity variations affect the average melt rate. In figure~\ref{fig:fig3}(a), we plot the normalized volume of ice $V(t)/V_0$ as a function of time for different $S_m$ with $\Delta S_v=5\rm\ g/kg$, $\Delta T=\SI{20}{K}$, and $H=\SI{5}{cm}$.
Interestingly, the melt rate shows a non-monotonic relation with the mean ambient salinity $S_m$, see figure~\ref{fig:fig3}(b). In that figure, to further quantify the melt rate, we have calculated the average melt rate $\bar{f}=1/t_{1/2}$, where $t_{1/2}$ represents the time needed to melt half of the initial volume, shown as dashed line in figure~\ref{fig:fig3}(a), and show $\bar{f}$ as a function of $S_m$ for various $\Delta S_v$. From the data points, one can see a non-monotonic dependence of $\bar{f}$ on $S_m$ observed for both two- and three-dimensional simulations: As $S_m$ increases, $\bar{f}$ first decreases and then increases, with a local minimum point depending on $\Delta S_v$. Note that we also tried different thresholds to calculate $\bar{f}$, which changes the absolute value of $\bar{f}$ while the trend remains the same.
The non-monotonic trend of melt rate is non-trivial; thus naturally the question is: Why is the dependence $f(S_m)$ non-monotonic?

\begin{figure}
\centering
\noindent\includegraphics[width=0.5\textwidth]{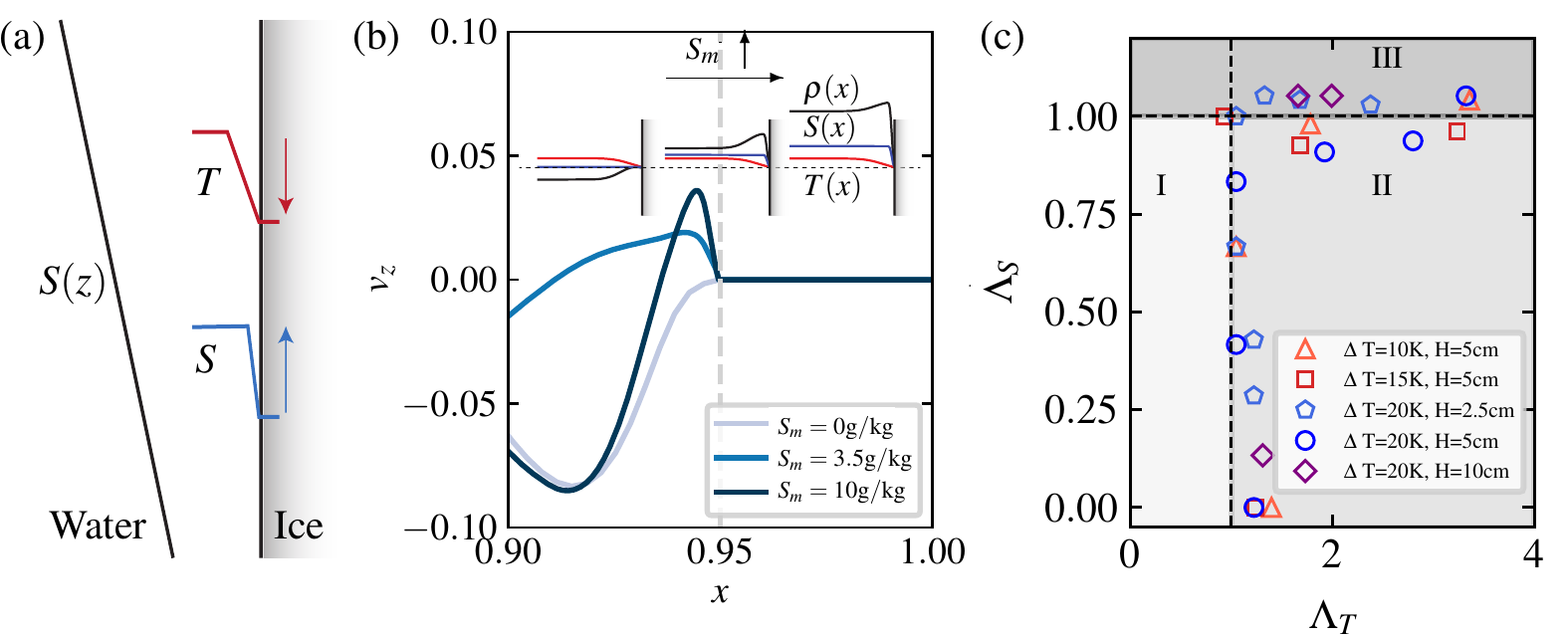}
\caption{
  (a) An illustration of the effect of temperature and salinity. The red and blue colors represent the buoyancy force driven by $T$ and $S$, with the arrows showing the direction of buoyancy. The black line represents the stable stratification.
  (b) The instantaneous vertical velocity (in free-fall velocity unit) profile at mid-height along $x$. $S_m=3.5~\rm{ g/kg}$ corresponds to the minimum melt rate. The inset figure illustrates the $T(x)$, $S(x)$, and $\rho(x)$ profiles at different ambient $S_m$. The dashed line shows the location of the melt front.
  (c) Location of the minimal melt rate in the parameter space spanned by $\Lambda_T$ and $\Lambda_S$ for various $\Delta T$ and $H$. The dashed lines show the prediction from eqs.~\ref{eq:model_ts} and \ref{eq:model_ss}. Regime I is `$T$-driven buoyancy', regime II is `$S$-driven buoyancy', and regime III is `stable stratification'.
  }
\label{fig:fig4}
\end{figure}

To understand this behaviour, we consider the energy balance in the system. A similar energy argument was adopted for the melting in fresh water \citep{yang2022abrupt}, where the density anomaly plays an important role. An illustration of the main energy terms driving the flow is shown in figure~\ref{fig:fig4}(a). The work for raising/sinking the fluid parcel in the stable stratification ($\Delta S_v$) is done by the buoyancy force, which is driven by both temperature $\Delta T$ and salinity $S_m$. When $S_m$ is small, temperature dominates the buoyancy, and the cold fresh meltwater moves downward. A circulation flow forms and transports cold water away and warm water towards the ice. When $S_m$ is large, salinity dominates the buoyancy, and the cold fresh meltwater moves upward, a circulation flow is also generated and melts the ice efficiently. However, at mediate $S_m$, the temperature- and salinity-induced buoyancy compensate with each other. In this case, fresh meltwater has almost the same density as the surroundings, which weakens the buoyancy-driven flow. The weakened flow near the melt front can be seen from figure~\ref{fig:fig4}(b), where we plot the vertical velocity profile in the horizontal direction for three different values of $S_m$. 

Quantitatively, when the stable stratification is weak, we consider the balance between the work done by thermal buoyancy to raise the fluid parcel over the domain $E_{\Delta T}=\frac{1}{2}C_bg\Delta T^2H$, and the work done by saline buoyancy $E_{{S_m}}=b_0gS_mH$, both based on the above given EOS eq.~\ref{eq:EOS}. We obtain
\begin{equation}
\frac{1}{2}C_bg\Delta T^2H=b_0gS_mH\ \rm{or}\ \Lambda_T=1,
\label{eq:model_ts}
\end{equation}
which means that the temperature and salinity-induced buoyancies compensate each other. 

When the stable stratification is strong (e.g. large $\Delta S_v$,  rightmost column of figure~\ref{fig:fig2}(a)), at low $S_m$ a layer of fresh water at the top emerges that has melted away faster. At large $S_m$ convection is stronger, also resulting in faster melting. Therefore, there is a minimum for medium $S_m$, which results from the competition between buoyancy (winning for low $S_m$) and stably stratification (winning at large $S_m$). We roughly estimated this minimum point by the balance between the potential energy induced by the saline stratification, which can be written as $E_{{\Delta S_v}}=N^2_0H^2=b_0g\Delta S_vH$ ($N_0$ is the buoyancy frequency), and the salinity-driven buoyancy. We obtain
\begin{equation}
b_0gS_mH=b_0g\Delta S_vH\ \rm{or}\ \Lambda_S=1.
\label{eq:model_ss}
\end{equation}
To check whether the results agree with the theory, we calculate the values of $\Lambda_T$ and $\Lambda_S$ corresponding to all minimum melt rate points. Then we plot them in figure~\ref{fig:fig4}(c) in the parameter space spanned by $\Lambda_T$ and $\Lambda_S$ for various $\Delta T$ and $H$. It can be seen that the data points for different $\Delta T$ and $H$ follow the same trend of $\Lambda_T=1$ (I-II) and $\Lambda_S=1$ (II-III), defining the transitions between the different regimes. 

\textbf{Conclusions and Outlook}: In summary, we have numerically studied ice melting in saline water, using direct numerical simulation with a realistic, nonlinear EOS. We have shown a non-monotonic dependence of the melt rate on the ambient salinity: as the ambient mean salinity increases, the melt rate first decreases and then increases. The physical origin of this non-monotonic dependence is the competition between thermally-driven buoyancy and salt-driven buoyancy, and the stable stratification due to the vertical salinity gradient. We derived a theoretical model based on an energy balance, which collapses the points of the minimum melt rate in the non-monotonic trend. Finally, we have shown the effect of salinity on the melt shape. Layered structures on the melt front have been observed, and the comparison of the layer thicknesses with the experimental results of \cite{huppert1980ice} shows quantitative agreement. 

From a broader perspective, our results show the ability of the phase field method to quantitatively model the melting process in multi-component turbulent flows \citep{hester2020improved}. Our numerical and theoretical results on the effect of salinity on the ice melt rate and shape can be applied to various saline water systems of geophysical relevance, e.g. sea ice, ice shelves, icy moons (which usually have even higher salinity than we explored), and more generally to multicomponent phase change materials.


\end{document}